\documentclass{article}

\usepackage{arxiv}

\usepackage[utf8]{inputenc} 
\usepackage[T1]{fontenc} 
\usepackage{hyperref} 
\usepackage{url} 
\usepackage{booktabs} 
\usepackage{amsfonts} 
\usepackage{nicefrac} 
\usepackage{microtype} 
\usepackage{lipsum}
\usepackage{amsmath}
\usepackage{graphicx}
\usepackage{makecell}
\usepackage{enumitem}
\usepackage{soul}
\usepackage{array,multirow}
\usepackage{lineno}
\usepackage[utf8]{inputenc}
\usepackage[english]{babel}
\usepackage{comment}
\usepackage[normalem]{ulem}
\usepackage{color}

\newcommand{\newtext}[1]{\textcolor{black}{#1}}

\title{\newtext{Design and Analysis of group-sequential clinical trials based on a modestly-weighted log-rank test in anticipation of a delayed separation of survival curves: A practical guidance}}

\author{
  Dominic Magirr \\
  Advanced Methodology and Data Science\\
  Novartis Pharma AG\\
  Basel, Switzerland\\
  \texttt{dominic.magirr@novartis.com} \\
   \And
  Jos\'e L. Jim\'enez \\
  Quantitative Safety and Epidemiology\\
  Novartis Pharma AG\\
  Basel, Switzerland\\
  \texttt{jose\_luis.jimenez@novartis.com} \\

}

\begin{document}
\maketitle

\begin{abstract}
A common feature of many recent trials evaluating the effects of immunotherapy on survival is that non-proportional hazards can be anticipated at the design stage. This raises the possibility to use a statistical method tailored towards testing the purported long-term benefit, rather than applying the more standard log-rank test and/or Cox model. Many such proposals have been made in recent years, but there remains a lack of practical guidance on implementation, particularly in the context of group-sequential designs. In this article, we aim to fill this gap. We discuss how the POPLAR trial, which compared immunotherapy versus chemotherapy in non-small-cell lung cancer, might have been re-designed to be more robust to the presence of a delayed effect. We then provide step-by-step instructions on how to analyse a hypothetical realisation of the trial, based on this new design. Basic theory on weighted log-rank tests and group-sequential methods is covered, and an accompanying \texttt{R} package (including vignette) is provided.
\end{abstract}

\section{Introduction}
For a homogeneous patient population, the primary analysis of a randomized controlled trial with a time-to-event endpoint is nothing more than a comparison of two cumulative distribution functions. Statistical analysis is made difficult, however, by right censoring, which precludes a simple comparison of means. The addition of one or more interim analyses complicates matters further. A standard solution is a group-sequential log-rank test, typically complimented with Kaplan-Meier estimates and a Cox proportional hazards model. Although successful in general, this strategy works less well for immuno-oncology trials, where the proportional hazards assumption is untenable. In this context, it is unlikely that the experimental drug will lead to an immediate improvement in survival. Rather, the survival curves are expected to be similar, or possibly favour the control arm, for a number of months, before diverging. The log-rank test, although valid, may have low power if the component of the test statistic corresponding to early timepoints is contributing noise without contributing signal. In addition, the estimated beta coefficient corresponding to the treatment term in the Cox model will no longer have a straightforward interpretation.

Numerous proposals have been made to replace the log-rank test with a weighted version that is tailored towards testing purported long-term improvements in survival \cite{harrington1982class,yang2010improved,gares2014comparison,karrison2016versatile,roychoudhury2021robust}. Uptake has been slow, however, in part due to concerns that such tests could produce counter-intuitive results when the hazard functions on the two arms cross \cite{freidlin2019methods}. To address such concerns, a "modestly-weighted" log-rank test has been proposed \cite{magirr2019modestly}, with the key property that if survival on the experimental drug is truly lower (or equal) to survival on control at all timepoints, then the probability of claiming a statistically significant improvement is less than $\alpha$. The modestly-weighted test also has considerably greater power than the standard log-rank test when there is a delayed treatment effect, as well as being straightforward to implement \cite{magirr2020non}.

In this paper, we aim to provide researchers with the guidance and tools necessary to use a modestly-weighted test in the context of a group-sequential design. Our emphasis will be on the practical side, since, from a methodological perspective, no new concepts are required. The modestly-weighted log-rank test belongs to the class of weighted log-rank statistic studied by Fleming \& Harrington \cite{fleming2011counting}, which, as shown by Tsiatis \cite{tsiatis1982repeated}, satisfy the standard independent increments assumption of  group-sequential theory \cite{jennison1999group, whitehead1997design}. We refer to Gillen \& Emerson   \cite{gillen2005information} for a detailed account of the methodology.

\section{Example: the POPLAR (NCT01903993) trial}
\label{sc_poplar}

We shall use the POPLAR trial \cite{fehrenbacher2016atezolizumab} as \newtext{a starting point for our discussion}. POPLAR was an open-label phase 2 randomized controlled trial of atezolizumab versus docetaxel for patients with previously-treated non-small-cell lung cancer. The key design assumptions, as well as a de-identified data set \cite{gandara2018blood}, are publicly available. The sample size was calculated assuming a median OS of 8 months for the control arm and a HR of 0.65, which translated into an assumed median OS of approximately 12.3 months for the atezolizumab arm, under an exponential model. Recruitment lasted 8 months. Three interim analyses were planned, with (two-sided) alpha levels of  0.0001, 0.0001, and 0.001. The final analysis of OS was performed when 173 deaths had occurred in the intention-to-treat (ITT) population, using a two-sided $\alpha$ level of 4.88\%. The trial enrolled a total of 287 patients.

\begin{figure*}[h]
  \centering
   \caption{Kaplan-Meier curves from the POPLAR trial.}
  \vspace{0.5cm}
  \includegraphics[scale=0.5]{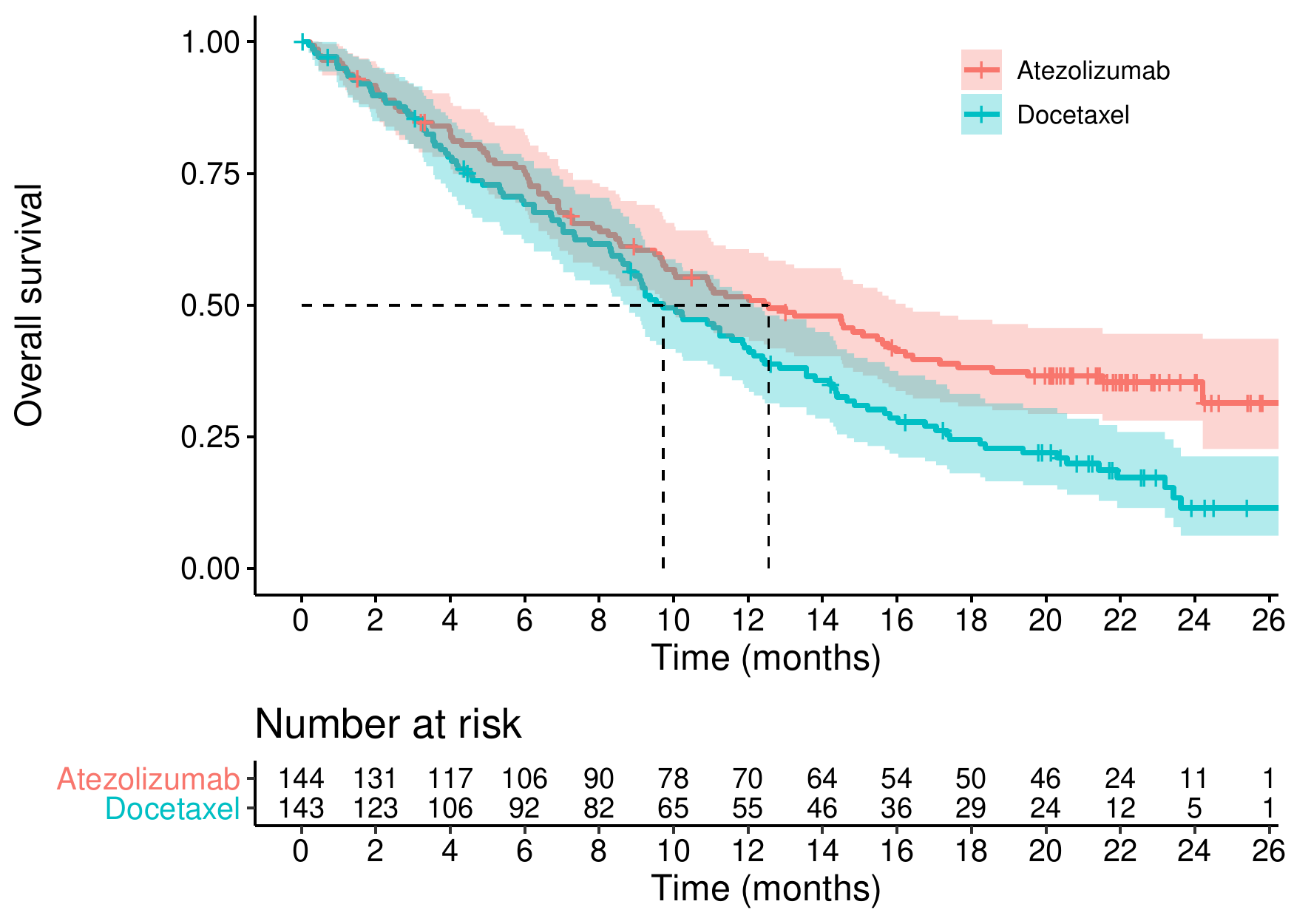}
  \label{figure_kaplan_meier_POPLAR}
\end{figure*}

A Kaplan-Meier estimate derived from the published data set \cite{gandara2018blood} is shown in Figure \ref{figure_kaplan_meier_POPLAR}. The curves display the typical late separation pattern often seen with immunotherapy agents. With the benefit of hindsight, but also based on observations from similar studies \cite{rahman2019deviation}, we will show how the trial might have been designed more robustly and efficiently, taking into account the potential for a delayed treatment effect.

\section{Methodology}
\label{sc_methodology}

\subsection{Weighted log-rank tests}

To perform a weighted log-rank test, we scan over the ordered event times $t_1,\ldots,t_k$, and take a weighted sum of the observed minus expected events on one of the treatment arms, where the expectation is taken assuming that the survival distributions on the two arms are identical. Let $n_{i,j}$ denote the number of patients at risk on treatment $i=0,1$ just prior to time $t_j$, and let $O_{i,j}$ denote the observed number of events on treatment $i=0,1$ at time $t_j$, with the expected number of events given by $E_{i,j} = O_{j}\times n_{i,j} / n_{j}$, where $n_j = n_{0,j}+n_{1,j}$ and $O_j = O_{0,j}+O_{1,j}$. Then the weighted log-rank test statistic is 
\begin{equation*}
    U_W := \sum_{j} w_j\left( O_{1,j} - E_{1,j}\right) \sim  N(0, V_W),
\end{equation*}
where
\begin{equation*}
    V_W = \sum_jw_j^2\frac{n_{0,j}n_{1,j}O_j(n_j - O_j)}{n_j^2(n_j - 1)}.
\end{equation*}
Intuitively, if the treatment is beneficial, we will tend to see fewer events on the experimental arm than would be expected assuming the curves are identical. We are hoping to see that $U_W << 0$, and, in particular, that the one-sided p-value, $p:=\Phi(U_W / \sqrt{V_W})$, is less than, e.g., $\alpha = 0.025$. Weights are pre-specified to boost the chances that $p<\alpha$, given the anticipated treatment effect.
The standard log-rank test uses $w_j = 1$, which is the most powerful choice under proportional hazards. Under a delayed-treatment-effect scenario, a popular alternative is the Fleming-Harrington-(0,1) test, which uses $w_j = 1 - \hat{S}(t_j-)$, where $\hat{S}(t_j-)$ is the Kaplan-Meier estimate of the pooled sample just prior to time $t_j$. Considerable care is necessary, however, since although the Fleming-Harrington-(0,1) test controls the type 1 error rate when survival curves are identical, it offers no guarantees regarding the direction of the effect \cite{magirr2020non, jimenez2019properties}. To put it another way: it offers a valid $\alpha$-level test when the null hypothesis is identical survival, $H_0:S_0(t)=S_1(t)$ for all $t$, but not when the null hypothesis is inferior (or identical) survival, $\tilde{H}_0:S_0(t)\leq S_1(t)$ for all $t$.  A safer choice that controls $\alpha$ also under $\tilde{H}_0$ is a  "modestly-weighted" log-rank test \cite{magirr2019modestly}, which uses $w_j = 1 / \max{\left\lbrace \hat{S}(t_j-), \hat{S}(t^*)\right\rbrace}$. Heuristically, the modestly-weighted test can be thought of as similar to an average landmark analysis from time $t^*$ to the end of follow up \cite{magirr2020non}. This interpretation is helpful at the design stage when pre-specifying $t^*$. \newtext{To be more explicit, if there are several candidate landmark times which might be of interest, then the earliest such timepoint may be a good candidate for $t^*$. More discussion on the choice of $t^*$ is given for a specific example below.}

\subsection{Group-sequential weighted log-rank tests}
\label{sc_group_sequential_wlr_tests}

For a group-sequential version of the weighted log-rank test, we must consider the joint  distribution of $U_W^{(1)},\ldots,U_W^{(K)}$, where $U_W^{(k)}$ denotes the test statistic at analysis $k$. As shown by Tsiatis \cite{tsiatis1982repeated}, asymptotically under $H_0$,

\begin{equation}\label{eqn_mvn}
    \left(\begin{array}{c}
         U_W^{(1)}  \\
         U_W^{(2)}  \\
          \vdots \\
           U_W^{(K)}  
    \end{array}\right) \sim N\left(\left( \begin{array}{c}
         0  \\
         0  \\
          \vdots \\
           0  
    \end{array}\right),
    \left( \begin{array}{cccc}
         V_W^{(1)} &  V_W^{(1)} & \cdots &  V_W^{(1)} \\
         V_W^{(1)} & V_W^{(2)} & \cdots& V_W^{(2)}  \\
         \vdots & \vdots& \ddots & \\
         V_W^{(1)} & V_W^{(2)} &&V_W^{(K)}  
    \end{array}\right)\right).
\end{equation}

A group-sequential test can be defined via the $K$ critical values, $c_1,\ldots,c_K$ such that 

\begin{equation}\label{eqn_c}
    p\left( \bigcap_{k\leq K} \left\lbrace \frac{U_W^{(k)}}{\sqrt{V_W^{(k)}}} > c_k\right\rbrace;~H_0 \right) = 1 - \alpha.
\end{equation}

There are many different ways to choose such critical values \cite{whitehead1997design, jennison1999group}. One flexible approach is to use a Hwang-Shih-DeCani alpha-spending function \cite{hwang1990group}. In this case, we must pre-specify an anticipated variance of the final test statistic, $\tilde{V}_{W}^{(K)}$. Then, at analysis $k$, for $k = 1,\ldots,K-1$, we find the cumulative alpha spend,

\begin{equation}\label{eqn_hsd}
   \alpha^*_k = \alpha \times\min\left\lbrace 1, \frac{1- \exp\left({-\gamma \sqrt{V_W^{(k)} / \tilde{V}_{W}^{(K)}}}\right)}{1 - \exp\left({-\gamma}\right)}\right\rbrace.
\end{equation}

Further defining $\alpha^*_K:=\alpha$, the critical value $c_k$ ($k = 1,\ldots,K$) is found via numerical integration, such that 

\begin{equation}\label{eqn_c_k}
    p\left( \bigcap_{l\leq k} \left\lbrace \frac{U_W^{(l)}}{\sqrt{V_W^{(l)}}} > c_l\right\rbrace ;~H_0\right) = 1 - \alpha^*_k.
\end{equation}

The parameter $\gamma$ can be chosen such that the stopping boundary resembles an O'Brien-Fleming boundary ($\gamma = -4$), a Pocock boundary ($\gamma = 1$), or something in between. 

\newtext{As an alternative to an alpha-spending function where the information fraction is given by the variance of the observed score statistic divided by the anticipated variance of the final score statistic, one might also consider to simply pre-fix $\alpha^*_1\leq \ldots\leq \alpha^*_K:=\alpha$. We shall evaluate both approaches in this paper. Note, however, that even more possibilities exist. For example, the information fraction could be based on the number of events, or calendar time, rather than the variance of the score statistic. In-depth discussion on alpha-spending functions can be found in \cite{jennison1999group}, for example.}

\section{Design}
\label{sc_design}

\subsection{Sample size calculation: fixed sample}

\begin{figure}[h]
  \centering
   \caption{Two potential alternative hypotheses for the POPLAR trial. \newtext{On the control arm, the survival distribution is exponential with median 8 months. For the no delay scenario, the survival distribution on the experimental arm is exponential with median 12.3 months. For the 4 month delay scenario, the survival distribution on the experimental arm is two-piece exponential with rate $\log(2)/8$ up to 4 months, and rate $\log(2)/16.6$ thereafter.}}
  \vspace{0.5cm}
  \includegraphics[scale=0.5]{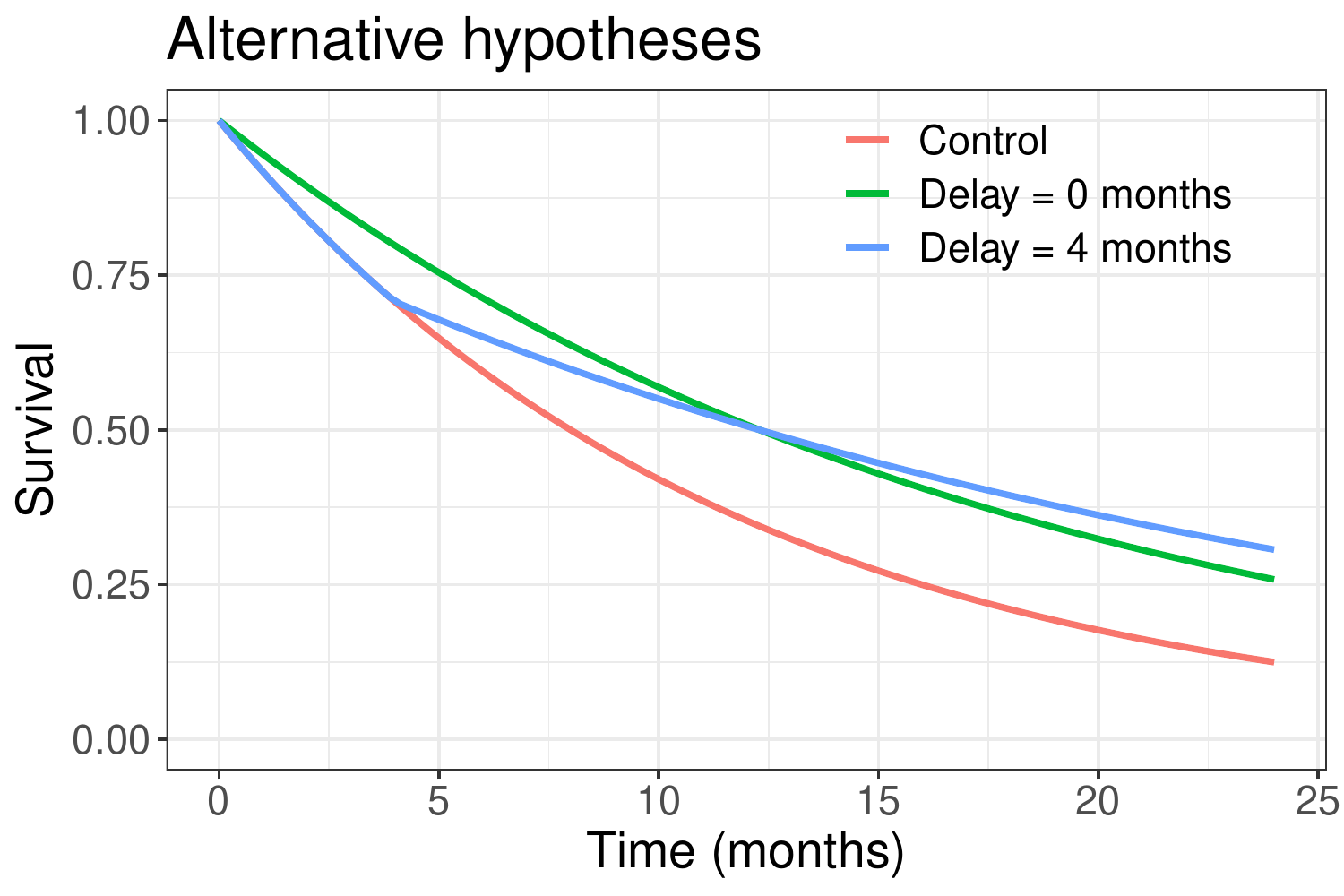}
  \label{figure_alternative}
\end{figure}

We now consider the alternative hypothesis, denoted by $H_1$. Figure \ref{figure_alternative} shows two potential alternative hypotheses that may have been considered for the POPLAR trial design. Our challenge is to find a design such that
\begin{equation}\label{eq_power}
    p\left(  U_W/\sqrt{V_W} < \Phi^{-1}(\alpha);~H_1  \right) = 1 - \beta.
\end{equation}
In time-to-event settings, power is driven by the number of events rather than the number of patients. The number of events is a function of the recruitment assumptions, time-to-event distributions, and the duration of follow up. Thus we have considerable flexibility, in theory at least, in how we design the trial to meet objective (\ref{eq_power}). If the sponsor of the study has large resources, it may be feasible to fix the duration of recruitment and follow-up to ensure that the study is completed in a timely manner. In this case, we adjust the recruitment rate, or, equivalently, the total number of patients, until (\ref{eq_power}) is satisfied. For example, the POPLAR trial specified 8 months of recruitment, plus a minimum follow-up time of 13 months, bringing the total trial duration to 21 months. Given these assumptions, as well as the time-to-event distributions in Figure \ref{figure_alternative}, the corresponding power of the standard log-rank test is shown in Table \ref{tab_samplesize_power_1} for a series of potential sample sizes. \newtext{The power has been calculated via numerical integration using the \texttt{R} package \texttt{gsdelayed} (available at \texttt{\href{https://github.com/dominicmagirr/gsdelayed}{github.com/dominicmagirr/gsdelayed}}), that we specifically developed to illustrate all the steps presented in this article. Computational details have already been described elsewhere \cite{magirr2019modestly}, but the basic idea is to calculate the expected number of events, as well as the expected average hazard ratio, based on the design assumptions, which can then be used as inputs to standard sample size formulae. The expected number of events at each analysis under the two design scenarios are also included in Table \ref{tab_samplesize_power_1}. Once a design has been chosen, one has the option during implementation to either keep the calendar time of the analyses fixed and allow the number of events to deviate from the plan, or instead to keep the number of events fixed and allow the calendar times of the analyses to deviate from the plan. In the implementation section below we shall opt for the latter approach}. 

\newtext{Table \ref{tab_samplesize_power_1} indicates} that under the proportional hazards ("0m delay") alternative, a sample size of 165 \newtext{patients} per arm would be sufficient to achieve 90\% power, \newtext{based on a trial duration of 21 months, with an expected number of events of 228}. However, under the non-proportional hazards ("4m delay") assumption, 180 \newtext{patients} per arm would be required \newtext{with an expected number of events of 244}. If, instead of the standard log-rank test, we use the modestly-weighted log-rank test with $t^*=6$, then the corresponding required \newtext{number of patients} per arm is 165 under proportional hazards and 150 under non-proportional hazards. \newtext{The corresponding expected number of events are 228 and 203, respectively.}  

\newtext{The choice of $t^*=6$ requires some explanation. Note first that $t^*=0$ is the same as the standard log-rank test, and for values of $t^*$ close to $0$ there will be little difference between these two tests. Also, as $t^* \rightarrow \infty$, the weights reduce to $w_j = 1 / \hat{S}(t_{j}-)$, which Gray \& Tsiatis \cite{gray1989linear} show are the optimal weights under a proportional-distributions cure-rate model. Loosely speaking, this is similar to comparing the survival probabilities at the end of follow-up. For intermediate values of $t^*$, the modestly-weighted log-rank test is similar to an average landmark analysis from time $t^*$ until the end of follow-up \cite{magirr2020non}. This means that if we have several different landmark times where we are potentially interested in the difference in survival curves, then the earliest such timepoint is a good candidate for $t^*$. Note, in particular, that if we anticipate a delay of 4 months, this does not imply that we should choose $t^*=4$. If we are confident about the delay then a somewhat later $t^*$ will have higher power. However, if there is some uncertainty regarding the delay, then choosing $t^*$ closer to zero protects the power in case proportional hazards does indeed hold. In addition to heuristic arguments, investigating operating characteristics for a range of $t^*$ is also helpful, as shown in our simulation study below.}

\begin{table}[h]
\centering
\caption{Relationship between number of patients per arm, expected total number of events, and power, using the standard log-rank test (LR) and the modestly weighted log-rank test (MWLR). Assuming uniform recruitment over 8 months, time-to-event distributions as given in Figure \ref{figure_alternative}, with analysis performed 21 months after the start of the trial.}
\begin{tabular}{|c|c|c|c|c|c|c|}
\hline
 & \multicolumn{2}{c|}{Total events}& \multicolumn{2}{c|}{Power LR} & \multicolumn{2}{c|}{Power MWLR ($t^* = 6$)} \\ 
\hline 
Patients per arm & 0m delay  & 4m delay  & 0m delay  & 4m delay  & 0m delay & 4m delay  \\
\hline
 150 & 207 & 203 & 0.87 & 0.84  & 0.87 & 0.91\\ 
 155 & 214 & 210 & 0.88 & 0.85 & 0.88 & 0.91\\ 
 160 & 221 & 217 & 0.89 & 0.86 & 0.89 & 0.92\\ 
 165 & 228 & 223 & 0.90 & 0.87 & 0.90 & 0.93\\ 
 170 & 235 & 230 & 0.91 & 0.88 & 0.90 & 0.94\\ 
 175 & 241 & 237 & 0.92 & 0.89 & 0.91 & 0.94\\ 
 180 & 248 & 244 & 0.92 & 0.90 & 0.92  &0.95\\ 
 \hline
\end{tabular}
\label{tab_samplesize_power_1}
\end{table}

To summarize, if we are confident in the 4-month delay assumption, and require 90\% power, then there is an approximate 20\% saving in \newtext{the number of patients per arm} from using the modestly-weighted log-rank test instead of a standard log-rank test. Even if we are not certain about the delayed effect, and would prefer to choose the sample size such that there is at least 90\% power under both proportional hazards and non-proportional hazards alternatives, there is still an approximate 10\%  reduction from using the modestly-weighted test.

\subsection{Adding an interim analysis (efficacy)}

We now consider adding an interim analysis for efficacy. Two choices are necessary: the timing of the interim analysis, and the amount of alpha to spend. In making these choices, we must consider our goal. For our example based on the POPLAR study, recruitment lasts 8 months, with a maximum trial length of 21 months. Unless the interim analysis is very early, all patients will have already been recruited, and most of the costs of the study will have already been incurred. The only incentive to stop early for efficacy is a reduction in the expected time until a decision. We could, for example, make choices that minimize the expected duration of the trial under the alternative hypothesis. Typically, however, there is a trade-off: the more we reduce the expected duration of the trial, the more we reduce the overall power. Or, if we decide to increase the maximum sample size to recover 90\% power, we must trade off a shorter expected duration versus a longer maximum duration. 

In Table \ref{tab_spending_functions}, expected duration and power are displayed for 10 potential designs. \newtext{Here, "two-stage" refers to a group-sequential design with one interim analysis, etc. Calculations are performed via numerical integration  using the \texttt{R} package \texttt{gsdelayed} as described above.} \newtext{The potential designs are merely a small selection of illustrative examples, but p}erhaps the three-stage design with interim analyses at 11 and 16 months stands out as an appealing option, based on a Hwang-Shih-DeCani spending function with $\gamma = -4$. This design reduces the expected duration of the study by 3.4 months with barely any reduction in power compared to a single-stage design. \newtext{In addition to power and expected duration under the alternative hypothesis,  many other metrics may be considered, such as the expected duration under the null hypothesis, or averaged over a range of scenarios. The topic is covered in-depth elsewhere \cite{jennison1999group, whitehead1997design}.}

\begin{table*}[t]
\centering
\caption{Expected duration and power of various design options. Based on a modestly-weighted log-rank test with $t^*=6$, sample size of 150 per arm, and uniform recruitment over 8 months. Survival distributions are assumed to follow the "4m delay" scenario in Figure \ref{figure_alternative}. }
\begin{tabular}{|c|c|c|c c c|ccc|}
\hline
&   & & \multicolumn{3}{c|}{E(months duration)} & \multicolumn{3}{c|}{Power} \\ 
\cline{4-9}
Design & Analysis times& Total events & $\gamma = -4$  & $\gamma = -1.5$  & $\gamma = 1$ & $\gamma = -4$ & $\gamma = -1.5$ & $\gamma = 1$\\
\hline
Single-stage & 21  & 203 & 21 & 21 &21 & 0.91 & 0.91 & 0.91\\ 
Two-stage & 11,~21 & 122,~203& 20.1 & 19.4 & 18.8 & 0.90 & 0.89 & 0.86\\ 
Two-stage & 16,~21 & 170,~203& 17.9 & 17.6 & 17.4 & 0.90 & 0.88 & 0.86\\ 
Three-stage & 11,~16,~21 &122,~170,~203 & 17.6 & 17.0 & 16.7 & 0.90 & 0.88 & 0.83\\ 
 \hline
\end{tabular}
\label{tab_spending_functions}
\end{table*}

\subsection{Adding an interim analysis (futility)}

Regulatory guidance generally steers towards futility stopping rules that are non-binding \cite{fda2019adaptive}. This means that we do not consider the futility stopping rule when we calculate the efficacy boundary to guarantee an $\alpha$-level test. If  non-binding futility rules are subsequently added, this has the effect of reducing both the type 1 error probability and the power.  

There are several ways that a futility rule could be specified \cite{gallo2014alternative}. We could, for example, consider a beta-spending function \cite{pampallona2001interim}. We could calculate the conditional power \cite{lachin2005review}, or the predictive power \cite{spiegelhalter1986monitoring}. Or we could specify a cut-off directly, either on the z-statistic scale or on the average-hazard-ratio scale. The latter has been implemented in \texttt{gsdelayed}. 

In the special case of time-to-event trials with an anticipated delayed effect, it should be recognised that a formal futility analysis may have limited value. As mentioned above, unless the interim analysis occurs very early, most patients will have been recruited, and most of the costs of the study already incurred. In addition, a stringent rule would risk stopping inappropriately before a late treatment effect has been given a chance to emerge.  This is not to say that the trial would never be stopped early. All such trials will be monitored by an independent data safety and monitoring board (DSMB). The DSMB will stop the trial promptly if the experimental drug is clearly harmful \cite{demets1994interim}.

\section{Implementation}
\label{sc_implementation}

\begin{figure}[h]
  \centering
   \caption{Switching from a study time perspective to an expected number of events perspective. Based on the alternative hypothesis (NPH)}
  \vspace{0.5cm}
  \includegraphics[scale=0.5]{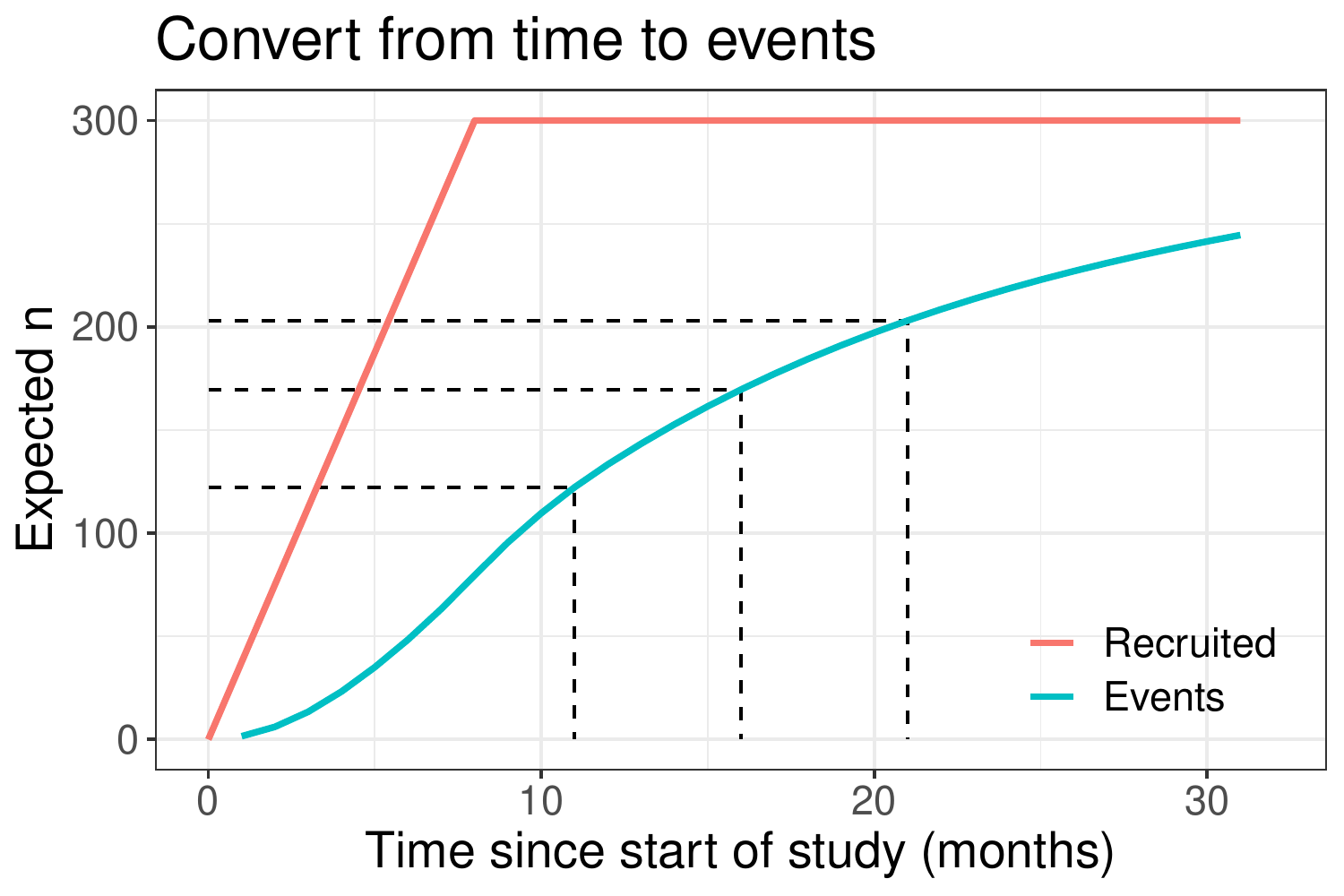}
  \label{figure_time_to_events}
\end{figure}

\begin{figure}[h]
  \centering
   \caption{Planned stopping boundaries with observed test statistics (modestly weighted log-rank test with $t^*=6$) overlaid (blue crosses). For illustration the z-statistic from the standard log-rank test is also included (red triangles).}
  \vspace{0.5cm}
  \includegraphics[scale=0.5]{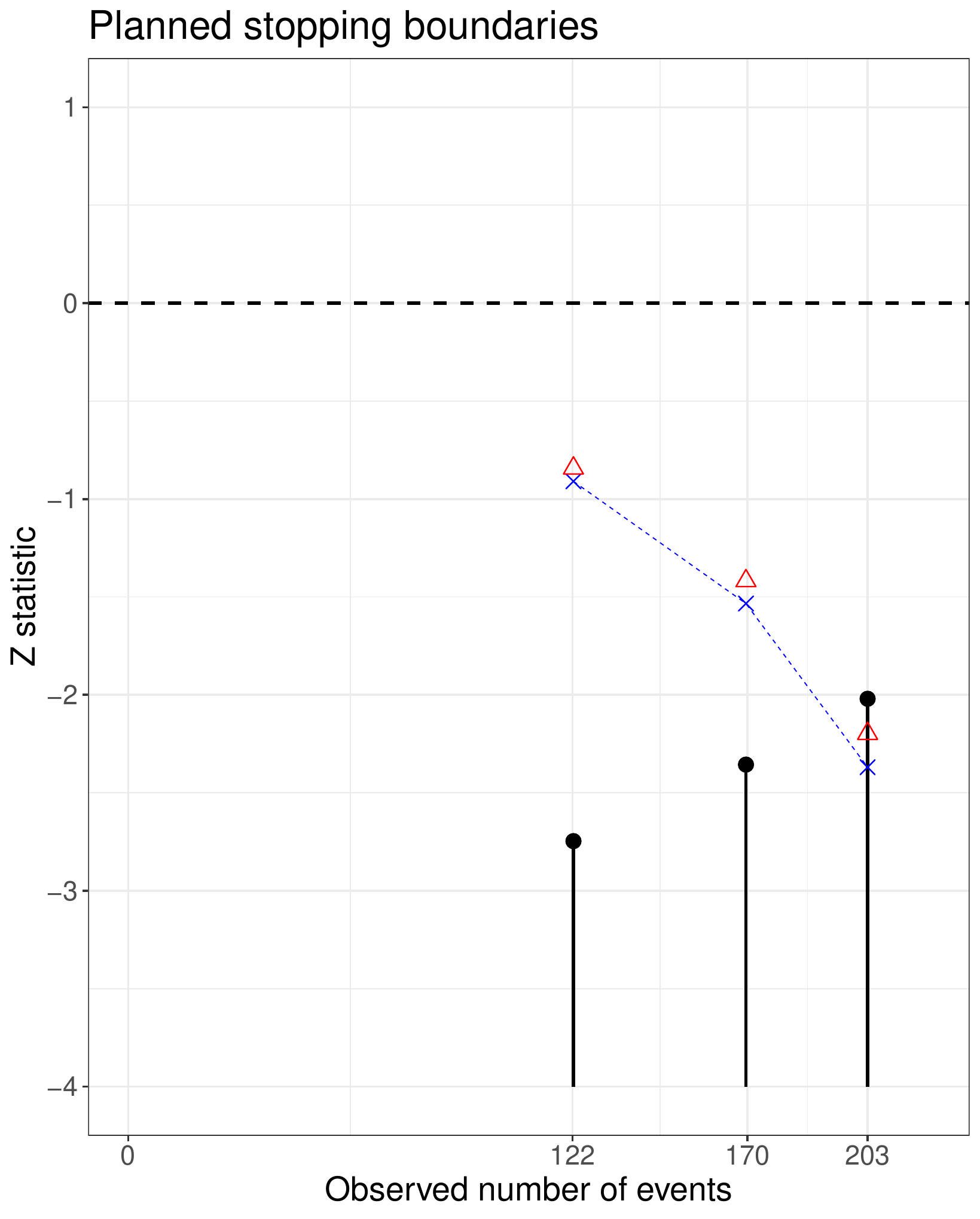}
  \label{figure_plan}
\end{figure}

We shall now walk through a hypothetical realization of the three-stage trial design from Table \ref{tab_spending_functions}. \newtext{We emphasize that this realization is not based on the results of the POPLAR study.} Figure \ref{figure_time_to_events} shows how the expected number of events corresponds to calendar time under the 4-month delay alternative. We see that the first interim, second interim and final analyses at months 11, 16 and 21, correspond to 122, 170 and 203 events, respectively. \newtext{As was described in Section \ref{sc_design}, we now have a choice of either fixing the calendar time points of the analyses, or fixing the number of observed events that will trigger each analyses. Here, we opt for the latter approach.} Having done so, the planned stopping boundaries are shown in Figure \ref{figure_plan}.

\newtext{Note, however, that the planned stopping boundaries are based on the assumed joint distribution of the test statistics under the design assumptions. Since the weights in the weighted log-rank test depend on the pooled survival distribution and recruitment distribution, which are inevitably misspecified at the design stage, we must update the boundary at each analysis in light of the observed variance of the score statistic. This is achieved via the alpha-spending function. Note that deviations from the exact numbers of planned events can also be handled in this way. As mentioned in Section \ref{sc_group_sequential_wlr_tests}, we shall consider two types of spending function. Firstly, we consider the Hwang-Shi-DeCani alpha-spending function with $\gamma = -4$, based on the information fraction $V_W^{(k)}/\tilde{V}_{W}^{(K)}$. With this approach, differences between the planned $\tilde{V}_W^{(k)}$ and observed $V_W^{(k)}$ lead to differences between the actual alpha spend at each analysis compared to the planned alpha spend. For this type of spending-function, the crucial parameter is the anticipated variance of the U statistic at the final analysis, denoted by $\tilde{V}_{W}^{(K)}$. For our example based on the design in the final row of Table \ref{tab_spending_functions}, we find via numerical integration that $\tilde{V}_{W}^{(3)} = 103.4$. The second type of alpha-spending approach that  we consider is to simply fix the cumulative alpha spend at each analysis as dictated by the design. In our case we have $\alpha_1^*= 0.00301$, $\alpha_2^* = 0.0106$, $\alpha_3^* = 0.025$. }

 \newtext{We now describe how our hypothetical trial proceeds:}

\begin{figure*}[h]
  \centering
   \caption{Kaplan-Meier curves at interim analyses 1 and 2, as well as at the final analysis.}
  \vspace{0.5cm}
  \includegraphics[scale=0.5]{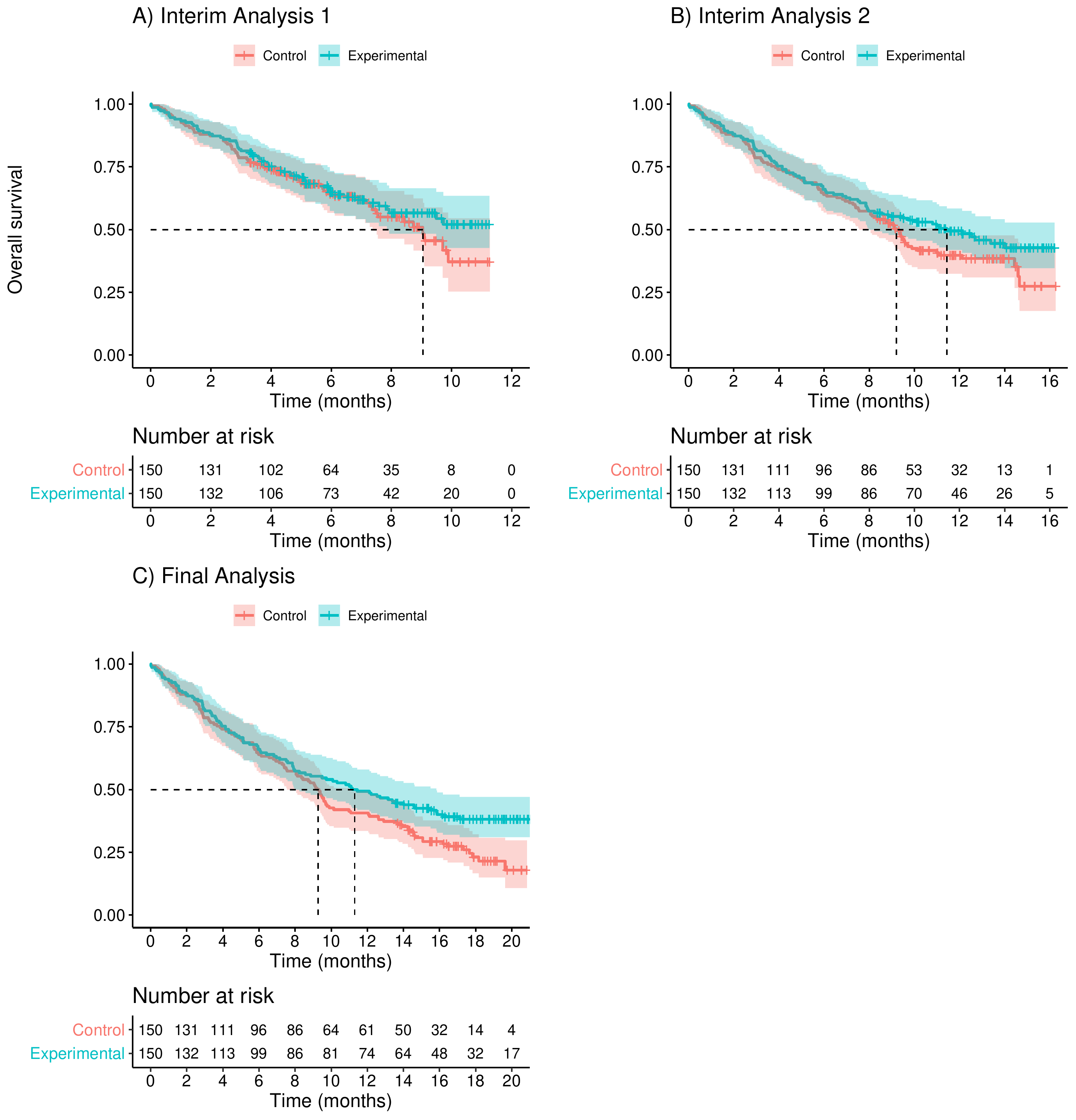}
  \label{figure_implementation_km}
\end{figure*}

\begin{itemize}

    \item Trial recruitment begins.
    
    \item \newtext{We conduct the first interim analysis after 122 events. Suppose we observe the data shown in Figure \ref{figure_implementation_km}A. Applying the modestly-weighted test, we find that $U_W^{(1)}=-6.46$ and $V_W^{(1)}=50.4$. The next step is to find the interim alpha spend. For the Hwang-Shi-DeCani alpha-spending function approach, we would plug the information fraction $t=V_W^{(1)}/\tilde{V}_{W}^{(3)}=0.487$ into (\ref{eqn_hsd}) to find $\alpha_1^* = 0.00281$. Or, using the fixed alpha-spending approach we would use $\alpha_1^* = 0.00301$. This corresponds to critical values on the Z-statistic scale of $-2.770$ or $-2.747$, respectively. In either case, since the observed Z-statistic $U_W^{(1)}/\sqrt{V_W^{(1)}} = -0.91$, the decision at the first interim would be to continue. This is represented graphically in Figure \ref{figure_plan} by the blue ``x'' at 122 events, where changes to the pre-planned critical value would be too small to be perceptible. }
    
    \item \newtext{We conduct the second interim analysis after 170 events. Suppose we observe the data shown in Figure \ref{figure_implementation_km}B. Again, applying the modestly-weighted test we find that $U_W^{(2)}=-13.6$ and $V_W^{(2)}=78.1$, so that $Z_2 = U_W^{(2)} / \sqrt{V_W^{(2)}}=-1.53$. The information fraction is now $V_W^{(2)}/\tilde{V}_{W}^{(3)}=0.755$ so that if we are using the Hwang-Shi-DeCani alpha-spending function we would have $\alpha_2^* = 0.0091$, whereas if we apply fixed alpha spending we would keep $\alpha_2^* = 0.0106$. Based on the observed correlation between $Z_1$ and $Z_2$, and using (\ref{eqn_c_k}), together with the critical value found at the first stage, this corresponds to critical values for $Z_2$ of $-2.42$ or $-2.35$, respectively. Since, in either case, $Z_2$ does not exceed the critical value, the decision would be to continue the trial. This is represented graphically in Figure \ref{figure_plan} by the blue ``x'' on top of 170 events. }
    
    \item \newtext{We conduct the final analysis after 203 events. Suppose we observe the data shown in Figure \ref{figure_implementation_km}C. Applying the modestly-weighted test, we find that $U_W^{(3)}=-23.4$ and $V_W^{(3)}=97.2$, so that $Z_3 = -2.37$. Notice in this case that $V_W^{(3)} < \tilde{V}_{W}^{(3)}$. With the pre-fixed alpha-spend we would have $\alpha^*_3 = 0.025$ and that would be the end of the trial. For the alpha-spending function with maximum information equal to $\tilde{V}_{W}^{(3)} = 103.4$, we could in principle carry on the trial to further analyses until the maximum information is reached. In  practice, it may be preferable to pre-specify a maximum of three analyses, and stipulate that all remaining $\alpha$ shall be spent at the final analysis. We shall assume this to be the case here, so that $\alpha^*_3 = 0.025$. Note that as a variation on this procedure, one could stipulate that if the maximum information has almost been reached, e.g., if $V_W^{(3)} / \tilde{V}_{W}^{(3)} > 0.95$, say, then the trial will be stopped with all remaining alpha to be spent, otherwise the trial will continue to a further analysis -- see the discussion in the simulation study section below.}
    \newtext{Based on the observed correlation between $Z_1$, $Z_2$ and $Z_3$, and using (\ref{eqn_c_k}), together with the critical values found at the first and second analyses, this corresponds to a critical value for $Z_3$ of $-2.00$ when applying the information-based spending fucntion, or $-2.01$ when applying the fixed alpha spending approach. Since, in either case, $Z_3$ exceeds the critical value, the final decision would be to reject the null hypothesis. This is represented graphically in Figure \ref{figure_plan} by the blue ``x'' on top of 203 events. } 
    
\end{itemize}

This hypothetical realization highlights the danger of a stringent futility analysis, as there is little separation along most of the Kaplan-Meier curves shown in Figure \ref{figure_implementation_km}A.

\newtext{In addition to testing the null hypothesis, some key tasks following completion of the trial are to calculate a p-value, and estimate the magnitude of treatment effect.} 

\newtext{In terms of defining a p-value based on a group sequential design, there is nothing special about our particular context involving a weighted log-rank test. We refer the reader to standard texts on group-sequential design theory \cite{whitehead1997design, jennison1999group}, and simply note  here that one }approach is a so-called "stage-wise ordering p-value", where earlier stops for efficacy are always considered more extreme evidence against the null hypothesis than later stops for efficacy. In the hypothetical trial described above, for the fixed alpha-spending approach, the stage-wise (one-sided) p-value would be:

\begin{equation}
\begin{split}
\label{eqn_msp}
& 1 - P_{H_0}\left( Z_1 >  -2.75~ \cap~ Z_2 > -2.35~ \cap ~Z_3 > -2.37\right) = 0.015
\end{split}
\end{equation}

In terms of treatment effect estimation, perhaps the most important tool is a Kaplan-Meier plot, which has the advantage of describing the entire survival curve. As has been noted by many authors \cite{magirr2020non,rufibach2019treatment,jimenez2020quantifying, roychoudhury2021robust}, in the setting of non-proportional hazards, there is no single-number summary measure that can adequately capture the full  information from the survival curves. Rather, it is considered helpful to report a range of single-number summary measures, including the difference in survival at fixed time points, differences in quantiles of the survival distributions, and differences in restricted mean survival times. 

In our hypothetical realization, we might focus on the survival probabilities at 18 months (0.38 on experimental versus 0.23 on control), the median survival times (11.3 versus 9.3 months), or the restricted mean survival times up to 18 months (11.2 versus 10.3 months).

\newtext{A further treatment effect measure that has been proposed in this context is a weighted hazard ratio, based on the weights used in the test statistic \cite{lin2017estimation}. This effect measure has the advantage that it is consistent with the null hypothesis test, in the sense that a confidence interval may be constructed that will exclude 1 if and whenever the null hypothesis is rejected. While this is an attractive feature, the interpretability of a weighted hazard ratio has also been debated \cite{bartlett2020hazards, lin2020rejoinder}}.

For all summary measures, a group sequential design introduces some bias, owing to the possibility to stop early on a random high. Various methods have been proposed that attempt to account for this bias \cite{pinheiro1997estimating, fan2004conditional,walter2019randomised}. They are rarely used in practice, however, with the justification often that the size of the bias is small, particularly if the interim analyses occur late \cite{freidlin2009stopping}. \newtext{Similarly, standard confidence intervals that ignore the group-sequential design are often presented. We refer the reader to \cite{robertson2021point} for in-depth discussion of this issue.}

\section{Robustness to design assumptions}

\begin{figure}[h]
  \centering
   \caption{Scenarios for the simulation study to assess robustness of methods to design assumptions. Survival on control arm is exponential with median either (A) 6 months, (B) 8 months, or (C) 8 months. On the experimental arm, the "Delay = 0 months" scenario is exponential with median (A) 9.2 months, (B) 12.3 months, or (C) 15.4 months. The "Delay = 4 months" scenario is two-piece exponential with a rate equal to the control arm rate until 4 months, and then a rate of (A) $\log(2) / 12.6$, (B) $\log(2) / 16.6$, or (C) $\log(2) / 21.6$, thereafter. The "Delay = 8 months" scenario is two-piece exponential with a rate equal to the control arm rate until 8 months, and then a rate of (A) $\log(2) / 25$, (B) $\log(2) / 35$, or (C) $\log(2) / 45$, thereafter. }
  \vspace{0.5cm}
  \includegraphics[scale=0.5]{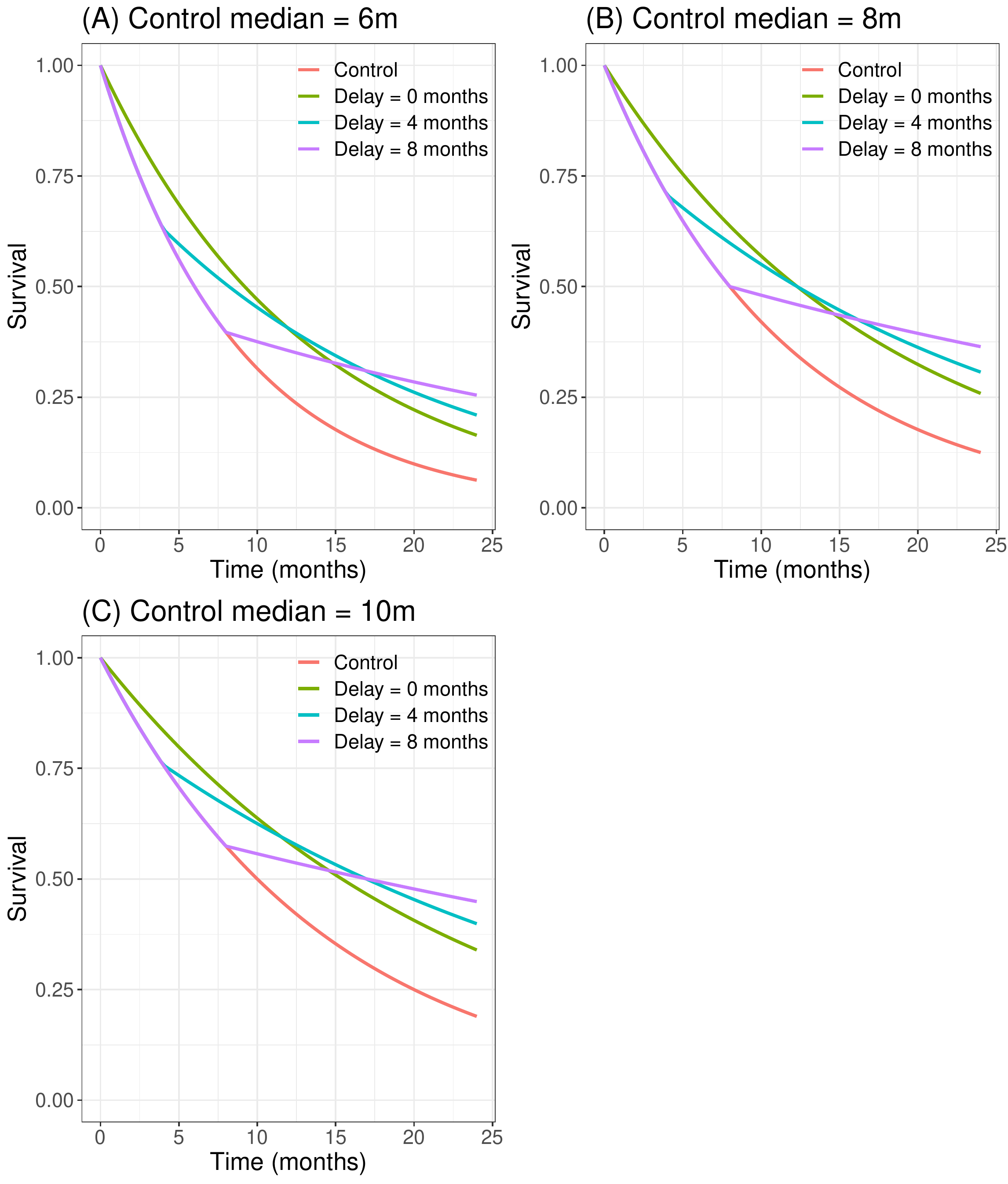}
  \label{figure_misspec}
\end{figure}

\newtext{Robustness of the proposed approach to design assumptions can be assessed via simulations. Here, as in Section \ref{sc_implementation}, we pre-specify a design that is based on a modestly-weighted log-rank test ($t^*=6$) with 150 patients per arm. First interim, second interim, and final analyses are triggered after 122, 170 and 203 events, respectively. According to the sample-size calculation in Section \ref{sc_design} (based on numerical integration), the power should be 90\% when recruitment is uniform over 8 months, and the survival distributions are as described in the "Delay = 4 months" scenario in Figure \ref{figure_misspec}(B). We now check this calculation via simulation, and assess how the power changes if the survival and/or recruitment distributions are misspecified as in Figure \ref{figure_misspec} (A),(B) and (C). Results are presented in the "$t^*=6$" columns of Table \ref{tab_sim_1}. When picking alternative scenarios, we have attempted to  offset longer delays with better long-term survival, but this does not necessarily imply that the scenarios represent equal magnitudes of clinical benefit. Note also that it is not the duration of the delay period per se that impacts power, but rather the proportion of the total number of events that occur during the delay period. In these simulations we have applied the alpha-spending function approach with the information fraction indexed by the variance of the score statistic. We have also stipulated that if 95\% of the maximum information is already available at the first interim analysis then the trial will stop with $\alpha_1^* = 0.025$. Similarly at the second interim if 97.5\% of the maximum information is already available. The trial stops after a maximum of three analyses with $\alpha_3^* = 0.025$. This procedure has the advantage of avoiding numerical instability, as well as mimicking more closely what would happen in practice. A further awkward issue is that, while unlikely, it is possible for the variance of the score statistics to decrease from one analysis to the next. At an interim analysis, the way that this is handled is to set the boundary to $-\infty$. At a final analysis we spend all remaining alpha in any case,  and correlations in the multivariate distribution (\ref{eqn_mvn}) are capped at 1.}

\newtext{Broadly speaking, we see that the operating characteristics are reasonably robust to misspecification of the control event rate, timing of the separation, and recruitment assumptions. The exception is when the control event rate is higher than expected and the delay is longer than expected. This situation is particularly challenging, with approximately 60\% of events occurring before there is any separation of the survival curves. If such a scenario is considered plausible at the design stage, one should consider choosing a larger value of $t^*$ and increasing the sample size.}

\newtext{Also included in Table \ref{tab_sim_1} are results from the standard log-rank test ($t^*=0$) and the modestly-weighted test with $t^* = 12$. These two tests have been applied with the same sample size, the same analysis trigger points, as well as the same alpha-spending function. The results follow the same pattern as was discussed for a single-stage design in Section \ref{sc_design}. For long delays, a large value of $t^*$ will increase power, but we need trade this off against some reduction in power under proportional hazards. }

\newtext{In Table \ref{tab_sim_2}, the simulation study has been repeated under the fixed alpha-spending approach. Conclusions are broadly the same, although the power appears somewhat more robust to model misspecification than when using the information-based spending function. This makes sense, given that the information-based spending function induces variation in the amount of interim alpha spend, whereas for an O'Brien-Fleming style boundary (as is used here) the fixed alpha-spending approach ensures that the interim alpha spend is kept low. }

\begin{table*}[t]
\centering
\caption{Simulation study with possibly misspecified survival and recruitment distributions. For each analysis method ($t^*=0,6,12$), the design is based on a control median of 8 months, a delay of 4 months, a sample size of 150 per arm, uniform recruitment over 8 months, analyses triggered after 122, 170 and 203 events, and a Hwang-Shi-DeCani ($\gamma = -4$) alpha-spending function. Recruitment scenario (1) is a uniform distribution over 8 months, as assumed in the design. In  scenario (2),  recruitment times $R$ are simulated under $P(R \leq r) = (r/15)^2$ for $r \leq 15$. See Figure \ref{figure_misspec} for full details of the survival distribution assumptions. Based on 10,000 simulations.}
\begin{tabular}{| c |c | c |c |c | c| c| c  | c |c |c |}
 \hline
\multirow{3}{*}{Recruitment}&\multirow{3}{*}{Treatment effect}& \multicolumn{9}{c|}{Power} \\
\cline{3-11}
&& \multicolumn{3}{c|}{Control median = 6m} &  \multicolumn{3}{c|}{Control median = 8m} & \multicolumn{3}{c|}{Control median = 10m}\\
 \cline{3-11}
 && $t^*=0$ &   $t^*=6$ &  $t^*=12$ & $t^*=0$ &   $t^*=6$ &  $t^*=12$ & $t^*=0$ &   $t^*=6$ &  $t^*=12$ \\
 \hline
 \hline
\multirow{4}{*}{1.}&Null & 0.028 & 0.024 & 0.027 & 0.027 & 0.025 & 0.028 & 0.023 & 0.026 & 0.024 \\
&Prop. Haz. & 0.85 & 0.84 & 0.82 & 0.85 & 0.86 & 0.84 & 0.85 & 0.86 & 0.85 \\
&Delay = 4m & 0.62 & 0.75 & 0.81 & 0.80 & 0.88 & 0.91 & 0.92 & 0.95 & 0.97 \\
&Delay = 8m & 0.18 & 0.22 & 0.39 & 0.66 & 0.77 & 0.89 & 0.93 & 0.96 & 0.99 \\
 \hline
\multirow{4}{*}{2.}& Null & 0.024 & 0.024 & 0.026 & 0.026 & 0.027 & 0.025 & 0.027 & 0.027 & 0.024 \\
&Prop. Haz. & 0.85 & 0.85 & 0.81 & 0.85 & 0.85 & 0.83 & 0.86 & 0.86 & 0.85 \\
&Delay = 4m & 0.62 & 0.74 & 0.81 & 0.80 & 0.88 & 0.91 & 0.93 & 0.95 & 0.97 \\
&Delay = 8m & 0.19 & 0.23 & 0.41 & 0.65 & 0.77 & 0.90 & 0.93 & 0.96 & 0.99 \\
 \hline
\end{tabular}
\label{tab_sim_1}
\end{table*}

\begin{table*}[t]
\centering
\caption{Simulation study with possibly misspecified survival and recruitment distributions. For each analysis method ($t^*=0,6,12$), the design is based on a control median of 8 months, a delay of 4 months, a sample size of 150 per arm, uniform recruitment over 8 months, analyses triggered after 122, 170 and 203 events, and a fixed alpha-spend per analysis, which is derived from a Hwang-Shi-DeCani ($\gamma = -4$) alpha-spending function applied to the design assumptions. Recruitment scenario (1) is a uniform distribution over 8 months, as assumed in the design. In  scenario (2),  recruitment times $R$ are simulated under $P(R \leq r) = (r/15)^2$ for $r \leq 15$. See Figure \ref{figure_misspec} for full details of the survival distribution assumptions. Based on 10,000 simulations.}
\begin{tabular}{| c |c | c |c |c | c| c| c  | c |c |c |}
 \hline
\multirow{3}{*}{Recruitment}&\multirow{3}{*}{Treatment effect}& \multicolumn{9}{c|}{Power} \\
\cline{3-11}
&& \multicolumn{3}{c|}{Control median = 6m} &  \multicolumn{3}{c|}{Control median = 8m} & \multicolumn{3}{c|}{Control median = 10m}\\
 \cline{3-11}
 && $t^*=0$ &   $t^*=6$ &  $t^*=12$ & $t^*=0$ &   $t^*=6$ &  $t^*=12$ & $t^*=0$ &   $t^*=6$ &  $t^*=12$ \\
 \hline
 \hline
\multirow{4}{*}{1.}&Null & 0.029 & 0.025 & 0.024 & 0.023 & 0.027 & 0.026 & 0.022 & 0.025 & 0.028 \\
&Prop. Haz. & 0.86 & 0.84 & 0.83 & 0.86 & 0.85 & 0.84 & 0.86 & 0.86 & 0.85 \\
&Delay = 4m & 0.61 & 0.77 & 0.82 & 0.80 & 0.88 & 0.91 & 0.92 & 0.95 & 0.97 \\
&Delay = 8m & 0.17 & 0.24 & 0.39 & 0.67 & 0.78 & 0.90 & 0.92 & 0.96 & 0.99 \\
 \hline
\multirow{4}{*}{2.}& Null & 0.026 & 0.026 & 0.026 & 0.025 & 0.027 & 0.024 & 0.023 & 0.026 & 0.027 \\
&Prop. Haz. & 0.85 & 0.84 & 0.82 & 0.85 & 0.85 & 0.84 & 0.86 & 0.85 & 0.85 \\
&Delay = 4m & 0.62 & 0.77 & 0.83 & 0.79 & 0.87 & 0.91 & 0.92 & 0.95 & 0.97 \\
&Delay = 8m & 0.20 & 0.27 & 0.42 & 0.66 & 0.77 & 0.89 & 0.92 & 0.96 & 0.99 \\
 \hline
\end{tabular}
\label{tab_sim_2}
\end{table*}

\section{Concluding remarks}
\label{sc_remarks}

Immunotherapy treatments often have delayed effects. We could use this knowledge to make phase 3 clinical trials more efficient, by focusing the test statistic on the purported long-term survival benefit, rather than using the standard log-rank test as a default. 

One potential barrier to realizing this increase in efficiency is a lack of guidance and software for implementing the more efficient methods in the context of group-sequential trials. In this paper, we have described in detail how to design and analyse a phase 3 trial in immuno-oncology using a group-sequential modestly-weighted log-rank test. We have also discussed the scope for a formal futility analysis in the special case of a time-to-event endpoint with an anticipated delayed effect. Lastly, we have illustrated how a range of single-number summary measures together help to quantify the treatment effect, which is important given that the hazard ratio lacks interpretability in this setting.

\section*{Data availability}

The data used to produce the Kaplan-Meier curves in Figure \ref{figure_kaplan_meier_POPLAR} is publicly available in \cite{gandara2018blood}. The \texttt{R} code used throughout the article is part of the package \texttt{gsdelayed}, which includes a vignette, and is available at \texttt{\href{https://github.com/dominicmagirr/gsdelayed}{github.com/dominicmagirr/gsdelayed}}.

\clearpage

\bibliographystyle{unsrt}
\bibliography{main}

\end{document}